\documentclass[graybox]{svmult}

\usepackage{mathptmx}       % selects Times Roman as basic font
\usepackage{helvet}         % selects Helvetica as sans-serif font
\usepackage{courier}        % selects Courier as typewriter font
\usepackage{type1cm}        % activate if the above 3 fonts are
\usepackage{makeidx}         % allows index generation
\usepackage{graphicx}        % standard LaTeX graphics tool
\usepackage{multicol}        % used for the two-column index
\usepackage[bottom]{footmisc}% places footnotes at page bottom

\usepackage{natbib}

  \def \teff {$T_{\mathrm{eff}}$}
  \def \logg {$\log g$}

\makeindex             % used for the subject index
\begin{document}

\title*{Characterization of Exoplanet-Host Stars}
\author{Vardan Adibekyan, S\'{e}rgio G.~Sousa and Nuno C.~Santos}
\institute{Vardan Adibekyan \at Instituto de Astrof\'isica e Ci\^encias do Espa\c{c}o, Universidade do Porto, CAUP, Rua das Estrelas, 4150-762 Porto, Portugal,\\ 
\email{vadibekyan@astro.up.pt}
\and S\'{e}rgio G.~Sousa \at Instituto de Astrof\'isica e Ci\^encias do Espa\c{c}o, Universidade do Porto, CAUP, Rua das Estrelas, 4150-762 Porto, Portugal,\\ 
\email{Sergio.Sousa@astro.up.pt}
\and Nuno C.~Santos \at Instituto de Astrof\'isica e Ci\^encias do Espa\c{c}o, Universidade do Porto, CAUP and
Departamento de F\'isica e Astronomia, Faculdade de Ci\^encias, Universidade do Porto, Rua das Estrelas, 4150-762 Porto, Portugal,\\ 
\email{nuno.santos@astro.up.pt}}

\maketitle

\abstract{Precise and, if possible, accurate characterization of exoplanets cannot be dissociated from the characterization of their host stars.
In this chapter we discuss different methods and techniques used to derive fundamental properties and atmospheric parameters of exoplanet-host stars. 
The main limitations, advantages and disadvantages, as well as corresponding typical measurement uncertainties of each method are presented.}

\section{Introduction}
\label{intro}

The discovery of the first\footnote{The detection of two terrestrial-mass companions around the pulsar PSR1257+12 had already been announced in 1992 \citep{Wolszczan92}.} extrasolar planet orbiting a main-sequence star, 51 Peg b \citep{Mayor95}, marks the start of observational exoplanetology. Exoplanet research experienced huge progress during the last two decades and has surely become a solid research field in contemporary astrophysics. Thanks to the fast progress in the development of instrumentation and observational techniques during the past decades, as of today (January 2017) there are more than 3500 planets detected, while several thousand candidates still await validation \citep{Coughlin-16}. 

Today, the main efforts in exoplanet research are moving towards the precise characterization of detected planets, 
including their statistical properties, as well as the detection of planets with progressively lower masses.
Despite the aforementioned progress, the study of extrasolar planets' properties via direct observations is still a very difficult task, 
and their precise study and characterization cannot be dissociated from the study of the host stars.
For example, we should be aware that transit measurements only provide us with the planet-to-star radius ratio, 
and the mass provided by radial-velocity measurements is dependent on the stellar mass. Thus, the characterization of
planets (e.g., mass, radius, density, and age) requires characterization of their hosts, and the accuracy of the planets' properties fundamentally 
depends on the achieved accuracy of the hosts' properties. 

It is very clear for the  exoplanet scientific community that poor characterization of planet hosts and planets themselves
is an important limitation, which cannot be always compensated even with large number statistics. A good example is the \textit{Kepler} mission, which 
provided thousands of stars with exoplanet candidates and an extremely large sample of stars with no 
detected planets that can be used for comparison analyses. However, the vast majority of these stars are poorly characterized, which obviously decelerates
the --- though still revolutionary --- fast advance in the field. The example of \textit{Kepler} and that of other ongoing \citep[e.g., \textit{Gaia}, K2;][]{Perryman-11,Howell-14} and upcoming \citep[e.g., \textit{TESS}, \textit{PLATO};][]{Ricker-14,Rauer-14} space missions motivated the community to start 
coordinating efforts to characterize the planet hosts. The importance of coordinating the exoplanet follow-up efforts has been addressed and intensively discussed in several recent meetings\footnote{For example, during ExoPAG 11 (\url{https://exoplanets.nasa.gov/exep/events/14/}) and the K2 meeting that took place in Porto in 2016 (\url{http://www.iastro.pt/research/conferences/k2meeting/}).}, and during the past few years several dedicated communities\footnote{SAG-14 (\url{https://exoplanets.nasa.gov/exep/exopag/sag/}).} and 
web interfaces\footnote{ExoFOP (\url{https://exofop.ipac.caltech.edu/}).} have been created with the goal of optimizing and coordinating the resources in 
exoplanet follow-up studies and characterization of their host stars.

Regarding the accuracy of the characterization of exoplanet hosts, many groups all over the world are intensively working on pushing down the precision limits and on developing methods that are less model-dependent and are most time-efficient. Unfortunately, direct measurements of physical properties of stars 
--- including exoplanet hosts --- are very rare and are possible for only specific targets.
The physical properties of the host stars are usually derived by using theoretical stellar evolutionary models and/or models of atmospheres. 
The uncertainties in the stellar model parameters can highly influence the final
accuracy with which properties of the stars and their planets  are measured \citep[e.g.,][]{Soderblom-10,Basu-12}. Asteroseismology is the tool that comes to help 
on improving our knowledge of fundamental properties of stars. It can provide properties for bright exoplanet-host stars (solar-type and red-giant stars, but not 
the cool dwarfs)
with very high accuracy. \textit{PLATO} will take full advantage of asteroseismic analyses to characterize all the planet hosts  brighter than 11th magnitude \citep{Rauer-14}.
Few dozen exoplanet hosts detected by \textit{TESS} will also benefit from asteroseismology \citep{Campante-16}. One should also not underestimate the importance of 
the \textit{Gaia} mission in the characterization of exoplanet hosts. Combined with ground-based, high-resolution spectroscopy, \textit{Gaia} will provide precise fundamental
properties (radii, luminosities, distances, and surface gravities) of exoplanet hosts.

We should also note that, in exceptional cases, planetary 
properties can be derived without using stellar models. For  example, the surface gravity of transiting exoplanets can be directly determined from the 
spectroscopic orbit of the parent star and the parameters measured directly from the transit light curve \citep{Southworth-07}. Absolute masses and radii
of planets can be also determined with very high precision (down to 1--2\%) for multi-planet systems --- with detectable gravitational interactions between planets --- when precise light curves of transits and radial-velocity (RV) data
are available \citep{Almenara-15}. Interestingly, it was proposed that the masses of transiting planets can be estimated based solely on the transmission spectrum 
\citep{deWit-13}.

It is very interesting to realize that not only knowledge about the host star helps to better understand the planet, but also sometimes observations of exoplanets
help characterizing the stars. For example, the stellar density can be directly derived from the transit light curve alone 
(\citealt{Seager-03}, but see \citealt{Kipping-14}), and spatially-resolved
stellar photospheres can be studied in detail when transiting planets are observed spectroscopically \citep[e.g.,][]{Collier-10, Cegla-16}.

In this manuscript we present the main methods and techniques that are widely used to characterize exoplanet-host stars. Together with the 
description of different methods and techniques we will also discuss the main limitations and achievable precision.

\section{Fundamental properties of stars: mass, radius and age}
\label{mass_radius}

The mass of RV-detected planets scales as $M^{2/3}$, where $M$ is the mass of the stellar host, while the radius of transiting planets is derived from the depth of the transit event and
the radius of the parent star. Since planet formation is a relatively fast process compared to the lifetime or age of most of the planet hosts,
stellar age can be used as a proxy for the age of planetary systems. Thus, basic characterization of exoplanets implies basic characterization of their hosts.

\runinhead{Stellar masses} Very precise dynamical masses can be derived for double-lined and single-lined (if the RV is derived for each component) eclipsing binaries, 
and for non-eclipsing double-lined spectroscopic binaries if astrometric orbits of the stars are known (usually through long-baseline interferometry). 
These techniques are quite well known \citep[for a recent comprehensive review, see][]{Torres-10} and can provide masses with an accuracy of better than 3\%. 

Unfortunately, direct determinations are usually impractical for most stars and indirect methods have to be used.
Different empirical and theoretical indirect methods are commonly used to determine the mass of single field stars. The stellar masses can be estimated by
using the spectroscopic surface gravity and luminosity of the star, provided the \teff \ is known. This method can give masses with a precision of 10--20\% depending on
the precision on $\log g$ and distance (parallax) of stars \citep{Sousa-11}. The masses can be empirically estimated by using mass-luminosity relations with a precision below 10\% \citep[e.g.,][]{Xia-10}. Empirical relations between stellar mass and stellar parameters (\teff, \logg, and [Fe/H]) by \citet{Torres-10} give
a scatter of only $\sim$6\% for main-sequence stars with masses above 0.6M$_{\odot}$. Finally, stellar masses can be determined by comparing stellar observed properties
with stellar evolutionary tracks \citep[e.g.,][]{Johnson-10, Sousa-15} or by using the power of asteroseismology \citep[e.g.,][]{Huber-12, Chaplin-14}. The latter method can lead to mass uncertainties below 5\% \citep[e.g.,][]{Chaplin-14}.

We should note that several studies suggested that the masses of planet-hosting evolved stars derived from evolutionary tracks can be largely overestimated 
\citep[e.g.,][]{Lloyd-11,Lloyd-13,Takeda-15}. However, \citet{Ghezzi-15} recently found very good agreement between
model-independent masses and the masses estimated using \textsc{parsec} evolutionary tracks \citep{Bressan-12} for a sample of 59 benchmark evolved stars.

\runinhead{Stellar radii} One of the most accurate ways of determining stellar radii is to measure the angular size of stars using interferometry. 
When precise distances (parallaxes) of these stars are known this method provides a practically direct measurement the radius that reaches 1--3\% precision on the angular diameter
\citep[e.g.,][]{Boyajian-13, Boyajian-14}. Until now, distances (parallaxes) of only nearby stars were known with high precision thanks to the \textit{Hipparcos} satellite. 
However, \textit{Gaia} will improve the situation, providing very precise distances for stars with much larger distances than \textit{Hipparcos} could reach.
We note that angular sizes and, consequently, linear radii of stars can also be determined using lunar occultations. This method has clear limitations (e.g.,
the Moon does not cover all the stars in the sky), but can provide radii with a precision of down to 3\% \citep[e.g.,][]{Richichi-97}.

Another direct technique to derive accurate stellar radii is to use double-lined eclipsing binary systems. 
The measured light curve and derived radial velocities of the two components can be used  to estimate the radii of the two stars with accuracies
of better than $\sim$1\% \citep[e.g.,][]{Lacy-05, Southworth-07}. A catalog of about 170 detached eclipsing binary systems with precise mass and radius measurements is presented in \citet{Southworth-15}.

Accurate direct measurements of stellar radii with interferometry and/or using eclipsing binaries can be used to develop empirical relations between 
radius and photometric colors \citep{Boyajian-14}, or else radius and stellar parameters \citep{Torres-10}. These empirical relations can be used to predict radii
of stars with errors less than 5\%.

Finally, distant stars are inaccessible for direct angular diameter measurements and so indirect methods are necessary to estimate their radii. 
Stellar radii can be derived from stellar evolution models by using the luminosity and effective temperature of the stars \citep[e.g.,][]{Santos-04, Torres-06} 
or, for trasiting systems,
by using the stellar density\footnote{Note that the derivation of this parameter also depends on the limb-darkening coefficient and orbital eccentricity of the 
transiting planet \citep[e.g.,][]{Mortier-13}.} (directly derived from the light curve) and \teff \ \citep[e.g.,][]{Sozzetti-07}. Radii of exoplanet hosts 
can also be derived by using asteroseismic quantities combined with \teff \ and stellar metallicity. This technique provides stellar radii with a typical precision 
of 2--4\% \citep[e.g.,][]{Chaplin-14}.

\runinhead{Stellar ages} Determination of accurate stellar ages is not an easy task, especially for field stars. Unlike stellar mass and radius, stellar age cannot
be directly measured and the use of stellar models is usually necessary to estimate ages. In exceptional cases, stellar ages can be determined 
without involving stellar models, namely, for young groups of stars through their kinematics \citep[e.g.,][]{Makarov-07} and for old metal-poor stars 
by using nucleocosmochronometry \citep[e.g.,][]{Ludwig-10}. 

The most common ways of estimating ages of exoplanet hosts are isochrone placement \citep[e.g.,][]{Pont-04, Takeda-07} and asteroseismology 
\citep[e.g.,][]{Aguirre-13, Campante-15}. We note that both methods require a knowledge of stellar atmospheric parameters.
Whereas the uncertainties on ages derived from stellar isochrones are typically not better than $\sim$20--30\%  
\citep[e.g.,][]{Jorgensen-05, Maxted-15}, asteroseismology can provide ages with a relative precision of about 10--20\% \citep[e.g.,][]{Kjeldsen-09, Aguirre-16}.

Alternatively, stellar ages can be derived by using empirical relations, calibrated  between age and rotation period \citep[e.g.,][]{Barnes-07}, age 
and chromospheric activity \citep[e.g.,][]{Lyra-05,Mamajek-08}, as well as age and chemical abundance ratios \citep[e.g.,][]{Nissen-15}. While these empirical
relations can provide relative high precision (depending on the calibration), their absolute values are difficult to establish.
For an excellent review on stellar age derivation with different techniques we refer the reader to \citet{Soderblom-10}.

\begin{svgraybox}
Summarizing, we can state that, when direct measurements are possible, masses and radii of individual stars can be derived with a precision of better than 1--3\%, whereas stellar ages can be estimated with an accuracy of better than a few percent.
For large numbers of  exoplanet hosts, stellar fundamental properties can be derived with a precision of $\sim$10--20\% for mass and radius,
and with a precision of 20--30\% for ages, assuming stellar atmospheric parameters are derived with high precision (see sections below).
\end{svgraybox}

\section{Stellar atmospheric parameters}
\label{stellar_param}

Accurate derivation of stellar atmospheric parameters (\teff, \logg, and metallicity/chemical abundances) is very important to fully characterize exoplanet-host stars. We need only to remember that the first interesting hint observed for exoplanet hosts was the correlation between giant-planet occurrence and 
stellar metallicity \citep[e.g.,][]{Gonzalez-97, Santos-01}, which had crucial importance for the advance of exoplanet formation theories.
For individual stars, direct measurements of stellar sizes and masses can be used to determine effective temperature and surface gravity 
without using stellar models. Stellar metallicity and chemical abundances of stars cannot be directly measured and stellar atmospheric models need to be used.

As discussed in the previous section, the direct determination of radii and masses, and hence \teff \ and \logg, is not possible for most stars. Hence, indirect methods
need to be used.  
Stellar atmospheric parameters (\teff, \logg, and [Fe/H] as a proxy for overall metallicity) can be derived with different methods and techniques. 
Photometric calibrations \cite[e.g.,][]{Onehag-09, Casagrande-10, Brown-11}, depending on
the photometric systems, can provide stellar parameters with reasonably high precision \citep{Smalley-14}. Profiles of individual lines 
\citep[e.g.,][]{Catanzaro-04, Catanzaro-13, Cayrel-11} and spectral line depth/equivalent width (EW) ratios \citep[e.g.,][]{Gray-91, Sousa-12} can also be used to determine 
different stellar parameters. Some of these methods can provide parameters with very high precision \citep[e.g., $\sim$2K in \teff;][]{Gray-97},
but with significantly less accuracy. Nevertheless, the most used and accurate techniques of deriving stellar parameters are provided by stellar spectroscopy.
For a comprehensive description of different methods for atmospheric parameter derivation we refer the reader to \citet{Gray-05} and \citet{Niemczura-14}.

The main spectral analysis techniques for the determination of stellar parameters can be divided into two main groups: the EW method
and the spectral synthesis method. In classical EW methods, measurements of EWs of isolated individual metallic lines are used to
derive stellar parameters assuming excitation equilibrium and ionization balance \citep[e.g.,][]{Santos-04a, Sousa-14}. Spectral synthesis methods yield 
stellar parameters by fitting the observed spectrum --- all, selected parts of the spectrum, or even a selection of lines --- with a synthetic one 
\citep[][]{Valenti-96, Malavolta-14}, with a library of pre-computed synthetic spectra \citep[][]{Recio-Blanco-06}, or a library of EWs \citep{Boeche-16}. 
Today there are many automatic tools designed to derive stellar parameters with the EW method \citep[e.g.,][]{Magrini-13,Tabernero-13,Sousa-14}, spectral 
synthesis techniques \citep[e.g.,][]{Prieto-06, Sbordone-14}, as well as tools that integrate different techniques, models of atmospheres and radiative transfer codes 
\citep{Blanco-Cuaresma-14}. For further details about these techniques we refer the reader to \citet{Niemczura-14} and \citet{Prieto-16}.

Both EW and spectral synthesis techniques have their advantages, disadvantages and limitations. The EW method is usually fast and relies on well selected lines. However,
this method  cannot be applied to fast-rotating stars or to stars with severe line-blended spectra. For these stars, spectral synthesis methods should be used. 
Synthesis techniques typically require more complicated computations for the generation of synthetic spectra and heavily depend on the line list and 
atomic/molecular data. A common limitation of spectroscopic methods is that they cannot constrain stellar surface gravity well 
\citep[e.g.,][]{Sozzetti-07, Mortier-13, Tsantaki-14}.
The impact of an unconstrained \logg \ on the derivation of other stellar parameters (\teff \ and [Fe/H]) is minimal for the EW-based curve-of-growth approach, 
while it has a significant impact for spectral-synthesis-based methods \citep{Torres-12, Mortier-13}. Luckily, surface gravity can be derived with high precision using 
asteroseismology \citep[e.g.,][]{Huber-13}, as well as for transiting systems from their light curves combined with spectroscopic \teff \ and metallicity \citep[e.g.,][]{Seager-03}.
From these two estimates, asteroseismic \logg 's are preferable, since transit-based \logg's might be less accurate when the eccentricity or the impact parameter of
the transiting planet is not well constrained \citep[][]{Huber-13}. \citet{Mortier-14} proposed an empirical correction --- based on the comparison 
of spectroscopic and asteroseismic \logg's --- for the spectroscopic surface gravity that depends only on the effective temperature. 
A word of caution  should be voiced here. It is not advisable to fix the surface gravity --- derived from other, non-spectroscopic method --- when
doing spectral analyses \citep[][]{Mortier-14,Smalley-14}. Fixing the \logg \ can bias the results and derivation of other atmospheric parameters.

The spectroscopic determination of stellar parameters  is affected by different factors, many of which are briefly discussed in \citet{Smalley-14}. The influence of
many of these factors (e.g., model atmosphere physics and input data) can be minimized when the spectral analysis is done in a homogeneous way. Consequently,
when homogeneous and high-quality data are used, an extremely high precision in stellar parameters can be achieved. For example, the latest works on solar twins that are 
based on differential line-by-line analysis report a precision (internal error) in atmospheric parameters of $\sim$10 K for \teff, $\sim$0.02 dex for \logg,
and $\sim$0.01 dex for [Fe/H] \citep[e.g.,][]{Ramirez-14, Adibekyan-16a}. However, one should note that when analyzing spectra of the same star obtained with
different instruments and at different epochs, dispersion of stellar parameters larger than the aforementioned precision can be obtained \citep[e.g.,][]{Bensby-14, Adibekyan-16a}.
Systematic errors, due to the model atmospheres, analysis method and atomic data are much larger than the random errors. 
Comparison of the results obtained with different methods for very large numbers of stars \citep[e.g.,][]{Bensby-14, Smiljanic-14}, as well as comparison of results with model-independent values for benchmark stars \citep{Jofre-14,Heiter-15} show that realistic typical errors 
on stellar parameters are not less than 50--100 K for \teff, 0.1--0.2 dex for \logg, and 0.05--0.1 dex for metallicity. Further discussion on the impact 
of using  different atmosphere models and different analysis strategies on the derivation of stellar parameters is presented in \citet{Lebzelter-12}.

\begin{svgraybox}
Homogeneous derivation of stellar parameters is crucial for characterizing exoplanet-host stars. The internal (relative) precision of atmospheric parameters 
can be as good as $\sim$10 K for \teff, $\sim$0.02 dex for \logg, and $\sim$0.01 dex for [Fe/H], but the overall precision of these parameters will be considerably smaller.
\end{svgraybox}

\section{Chemical abundances of exoplanet-host stars}
\label{abundanes}

Exoplanet-related research always requires high precision and accuracy.
If very high-precision measurements are needed to detect planets, likewise, finding possible abundance
differences between stars with and without planets also requires accurate and homogeneous abundance determinations.
Many studies aimed at clarifying whether the planet-hosting stars  are different from stars
without planets in their content of individual heavy elements other than iron 
\citep[e.g.,][]{Fischer-05, Robinson-06, DelgadoMena-10, Adibekyan-12a, Adibekyan-15, Suarez-16}. In particular, it was shown that
metal-poor hosts tend to show systematic enhancement in $\alpha$ elements \citep[][]{Haywood-09, Adibekyan-12b, Adibekyan-12c}.
Accurate knowledge of abundances of individual heavy elements and specific elemental ratios (e.g., Mg/Si and C/O) in stars with planets are also very 
important because they are expected to control the structure and composition of terrestrial planets \citep[e.g.,][]{Grasset-09, Thiabaud-14, Dorn-15}.

Once the atmospheric parameters of stars are known, chemical abundances of individual elements can be derived spectroscopically by  
EW or spectral synthesis techniques. Many research groups are intensively working on the derivation of chemical abundances
in stellar atmospheres of stars with and without planets. The derivation of chemical abundances may seem very trivial, however, a simple comparison of the 
(discrepant) results obtained for the same elements from the same data in the same stars, but with different methods, shows that there are important factors
(e.g., line list and atomic data, continuum normalization, hyperfine structure, damping, microturbulence, NLTE effects, atmospheric model)
that need to be deeply investigated. Intensive and comprehensive discussion about the possible issues can be found in several recent articles 
\citep[e.g.,][]{Smiljanic-14, Jofre-15, Hinkel-16} that had as a common goal to \textit{open the black box of stellar element abundance determination} \citep{Jofre-16}.

As for the stellar parameters, when studying solar twins and solar analogs (i.e., stars that are very similar to our Sun in terms of stellar parameters)
extremely precise --- accuracy still can be an issue --- chemical abundances at the level of $\sim$0.01 dex can be obtained 
\citep[e.g.,][]{Ramirez-10, GH-13,Adibekyan-16b,Saffe-16}. High-precision abundances (at the level of $\sim$0.05--0.10 dex) can be also obtained 
for large samples of cool stars if high-quality data are used and, importantly, if the spectral analysis is done in a homogeneous way 
\citep[e.g.,][]{Adibekyan-12a, Bensby-14, Adibekyan-15a, Mikolaitis-16}. However, if the data are compiled from different sources, or different methods 
were used to derive abundances, then the results should be taken with caution. Method-to-method or study-to-study dispersion of chemical abundances
can be larger than 0.10--0.20 dex \citep[e.g.,][]{Hinkel-14, Smiljanic-14}.

\begin{svgraybox}
As for the atmospheric parameters, homogeneous derivation of chemical abundances is important to achieve high precision. 
Elemental abundances for large samples of cool stars can be derived with a typical internal (relative) precision of $\sim$0.05 dex, but the accuracy of 
these derivations will be smaller.
\end{svgraybox}

\section{Other properties of exoplanet-host stars}
\label{other}

\runinhead{Kinematics} Kinematics, or Galactic space-velocity components of stars, can be computed when a star's proper motion, radial velocity and parallax are known
\citep[e.g.,][]{Johnson-87}.
The kinematics of exoplanet-host stars and their relation to different
stellar populations and moving groups have been discussed in several  works \citep[e.g.,][]{Barbieri-02, Reid-02,Ecuvillon-07, Adibekyan-12c, Gaidos-17}.
Most papers have not reported any significant kinematic peculiarity of planet-hosting stars \citep[e.g.,][]{Gonzalez-99, Barbieri-02}.
Conversely, \citet{Haywood-08, Haywood-09}, combining the chemical and kinematic properties of exoplanet hosts, concluded that most metal-rich 
stars that host giant planets originate from the inner Galactic disk. The same scenario for the origin of metal-rich planet hosts is explored in a few other
works \citep[e.g.,][]{Ecuvillon-07, Santos-08, Adibekyan-14}.

\runinhead{Activity} Understanding stellar magnetic activity phenomena (such as spots, faculae,
plages) is very important for different fields of stellar physics and exoplanetary science, as well as for planetary climate
studies. Studying magnetic activity in stars of different stellar parameters and activity levels provides an
opportunity for detailed tests of stellar/solar dynamo models. From the exoplanetary side, it is well known
that stellar active regions, combined with the stellar rotation, can induce signals in high-precision photometric
and radial-velocity observations. These activity-induced signals may lead to masking or mimicking of exoplanet
signals \citep[e.g.,][]{Queloz-01, Dumusque-12, Oshagh-13, Santos-14}. Moreover, these signals constitute one of the
main obstacles to the detection and precise characterization of low-mass/small-radius planets, the
major goal of future instruments. Several indices (re-emission in the Ca II H \& K lines, Mg II h \& k lines, Ca
infrared triplet, Na I D doublet, H$\alpha$) exist to characterize the activity of stars 
\citep[e.g.,][]{Baliunas-95, Kurster-03, Mamajek-08, GomesdaSilva-11, Haswell-12, Mathur-14}. 
The applicability of these indices is restricted, as it depends on the spectral type of the stars and spectral coverage of the used spectrograph.
The dependence of stellar activity on the planet-star interaction was discussed in several observational 
and theoretical studies \citep[e.g.,][and references therein]{Figueira-16} yielding contradictory results. 

\runinhead{Rotation} The most common ways of measuring stellar rotation are through spectroscopy \citep[e.g.,][]{Benz-81, Donati-97} and photometry 
\citep[e.g.,][]{Irwin-09, McQuillan-13}. These techniques --- depending on the quality of the data and properties of the stars --- can provide rotation velocities\footnote{
Note that spectroscopy usually provides the $v\sin i$ and photometry gives the rotational period of the stars.}
with a precision of better than $\sim$10\% \citep[for the limitations and advantages of either technique, see][]{Bouvier-13}. Recent studies show
that the stars with planets (or with planet candidates in the case of \textit{Kepler}) rotate more slowly than stars without known planets 
\citep[e.g.,][]{Takeda-10, Gonzalez-15}. Moreover, it appears that only slow-rotating stars host close-in planets. The slow rotation of exoplanet host stars --- if not a selection and/or detection bias --- can be caused by early star-disk interactions \citep{Bouvier-08}.

\section{Conclusion}
\label{conclusion}

Precise and accurate characterization of exoplanet-host stars is crucial to the detailed investigation of exoplanets themselves.
Moreover, precise determination of stellar parameters is important to study the star-planet connection.
There are different ways of characterizing stars with and without planets. Some of these methods are independent of stellar models, hence fundamental, although most
are not. The combination of different methods can provide precise, and even accurate, stellar parameters and chemical abundances of exoplanet hosts.

When studying statistical properties of exoplanets or of their hosts it is very important to use information (parameters) as homogeneous as possible. 
A catalog of exoplanet hosts with stellar parameters derived and compiled in a homogeneous way is presented in \citet{Santos-13} and, for transiting systems, 
in \citet[][]{Southworth-12}.

\begin{acknowledgement}
VA would like to thank the SOC for inviting him to the \textit{IV$^{th}$ Azores International Advanced School in Space Sciences} 
held in the Azores Islands, Portugal.
VA, NCS and SGS acknowledge the support from Funda\c{c}\~ao para a Ci\^encia e a Tecnologia (FCT) through national funds
and from FEDER through COMPETE2020 by the following grants UID/FIS/04434/2013 \& POCI-01-0145-FEDER-007672, 
PTDC/FIS-AST/7073/2014 \& POCI-01-0145-FEDER-016880, and PTDC/FIS-AST/1526/2014 \& POCI-01-0145-FEDER-016886. 
VA, NCS and SGS acknowledge the support from FCT through Investigador FCT contracts IF/00650/2015/CP1273/CT0001, IF/00169/2012/CP0150/CT0002,
and IF/00028/2014/CP1215/CT0002. 
\end{acknowledgement}
\bibliographystyle{apj}
\bibliography{references}
\end{document}